\documentclass[runningheads]{cl2emult}
\usepackage{epsfig}
\def\cal{\mathcal}

\begin{document}

\title*{The Evolutionary Unfolding of Complexity}
\author{James P. Crutchfield \and Erik van Nimwegen}
\institute{Santa Fe Institute, 1399 Hyde Park Road, Santa
Fe, New Mexico 87501. \{chaos,erik\}@santafe.edu.}

\date{draft \today}

\maketitle

% ************************* ABSTRACT *************************

\begin{abstract}
  
We analyze the population dynamics of a broad class of fitness
functions that exhibit epochal evolution---a dynamical behavior,
commonly observed in both natural and artificial evolutionary
processes, in which long periods of stasis in an evolving population
are punctuated by sudden bursts of change. Our
approach---{\em statistical dynamics}---combines methods from both
statistical mechanics and dynamical systems theory in a way that offers
an alternative to current ``landscape'' models of evolutionary
optimization. We describe the population dynamics on the macroscopic
level of fitness classes or phenotype subbasins, while averaging
out the genotypic variation that is consistent with a macroscopic state.
Metastability in epochal evolution occurs solely at the macroscopic
level of the fitness distribution. While a balance between selection
and mutation maintains a quasistationary distribution of fitness,
individuals diffuse randomly through selectively neutral subbasins in
genotype space.  Sudden innovations occur when, through this diffusion,
a genotypic portal is discovered that connects to a new subbasin
of higher fitness genotypes. In this way, we identify innovations
with the unfolding and stabilization of a new dimension in the
macroscopic state space.  The architectural view of subbasins and
portals in genotype space clarifies how frozen accidents and
the resulting phenotypic constraints guide the evolution to higher
complexity.


\begin{center}
Santa Fe Institute Working Paper 99-02-015
\end{center}
\noindent
To appear in {\bf Evolution as Computation}, L. F. Landweber, E.
Winfree, R. Lipton, and S. Freeland, editors, Lecture Notes in
Computer Science, Springer-Verlag (1999). Proceedings of a DIMACS
Workshop, 11-12 January 1999, Princeton University.
\end{abstract}

% ****************************************************************

%\begin{multicols}{2}

%\tableofcontents

% ************************* INTRODUCTION *************************

\section{Evolutionary Computation Theory}

The recent mixing of evolutionary biology and theoretical computer
science has resulted in the phrase ``evolutionary computation'' taking
on a variety of related but clearly distinct meanings.

In one view of evolutionary computation we ask whether
Neo-Darwinian evolution can be productively analyzed in terms of how
biological information is stored, transmitted, and manipulated. That
is, Is it helpful to see the evolutionary process as a computation?

Instead of regarding evolution itself as a computation, one might ask
if evolution has produced organisms whose internal architecture and
dynamics are capable {\em in principle} of supporting arbitrarily
complex computations. Landweber and Kari argue that, yes, the
information processing embedded in the reassembly of fragmented gene
components by unicellular organisms is quite sophisticated; perhaps
these organisms are even capable of universal computation
\cite{Land99a}. It would appear, then, that evolved systems themselves
must be analyzed from a computational point of view.

Alternatively, from an engineering view we can ask, Does Neo-Darwinian
evolution suggest new approaches to solving computationally difficult
problems? This question drives much recent work in evolutionary
search---a class of stochastic optimization algorithms, loosely based
on processes believed to operate in biological evolution, that have been
applied successfully to a variety of different problems; see, for
example, Refs.
\cite{Back96a,ICGA91,Cham95a,Davis91a,ICGA95,ICGA93,Goldberg89c,Koza93a,Mitchell96a}
and references therein.

Naturally enough, there is a middle ground between the scientific
desire to understand how evolution works and the engineering desire
to use nature for human gain. If evolutionary processes do embed
various kinds of computation, then one can ask, Is this biological
information processing of use to us? That is, can we use biological
nature herself to perform computations that are of interest to us? A
partial, but affirmative answer was provided by Adelman, who mapped the
combinatorial problem of Directed Hamiltonian Paths onto a
macromolecular system that could be manipulated to solve this well
known hard problem \cite{Adel94a}.

Whether we are interested in this middle ground or adopt a scientific
or an engineering view, one still needs a mathematical framework with
which to analyze how a population of individuals (or of candidate
solutions) compete through replication and so, possibly, improve
through natural (or artificial) selection. This type of evolutionary
process is easy to describe. In the Neo-Darwinian view each individual
is specified by a {\em genotype} and replicates (i) according to its
{\em fitness} and (ii) subject to genetic variation. During the
passage from the population at one {\em generation} to the next, an
individual is translated from its genotypic specification into a form,
the {\em phenotype}, that can be directly evaluated for fitness and so
selected for inclusion in the next generation. Despite the ease of
describing the process qualitatively, the mechanisms constraining and
driving the population dynamics of evolutionary adaptation are not
well understood.

In mathematical terms, evolution is described as a nonlinear
population-based stochastic dynamical system. The complicated dynamics
exhibited by such systems has been appreciated for decades in the
field of mathematical population genetics \cite{Hartl&Clark}. For
example, the effects on evolutionary behavior of the rate of genetic
variation, the population size, and the genotype-to-fitness mapping
typically cannot be analyzed separately; there are strong, nonlinear
interactions between them. These complications make an empirical
approach to the question of whether and how to use evolutionary
optimization in engineering problematic.  They also make it difficult
to identify the mechanisms that drive behavior observed in
evolutionary experiments. In any case, one would like to start with
the basic equations of motion describing the evolutionary process, as
outlined in the previous paragraph, and then predict observable
features---such as, the time to find an optimal individual---or, at a
minimum, identify mechanisms that constrain and guide an evolving
population.

Here we review our recent results that address these and similar
questions about evolutionary dynamics. Our approach derives from an
attempt to unify and extend theoretical work that has been done in the
areas of evolutionary search theory, molecular evolution theory, and
mathematical population genetics. The eventual goal is to obtain a
more general and quantitative understanding of the emergent mechanisms
that control the population dynamics of evolutionary adaptation and
that govern other population-based dynamical systems.

\section{Epochal Evolution}

To date we have focused on a class of population-dynamical behavior
that we refer to as {\em epochal evolution}. In epochal evolution,
long periods of stasis ({\em epochs}) in the average fitness of the
population are punctuated by rapid {\em innovations} to higher
fitness.  These innovations typically reflect an increase of
complexity---that is, the appearance of new structures or novel
functions at the level of the phenotype. One central question then is,
How does epochal evolutionary population dynamics facilitate or impede
the emergence of such complexity?

Engineering issues aside, there is a compelling biological motivation
for a focus on epochal dynamics. There is the common occurrence in
natural evolutionary systems of ``punctuated equilibria''---a process
first introduced to describe sudden morphological changes in the
paleontological record \cite{Gould&Eldredge77}. Similar behavior has
been recently observed experimentally in bacterial colonies
\cite{Elena&Cooper&Lenski96} and in simulations of the evolution of
t-RNA secondary structures \cite{Font98a}. This class of behavior
appears sufficiently general that it occurs in artificial evolutionary
systems, such as evolving cellular automata
\cite{Crutchfield&Mitchell95a,MitchellEtAl93c} and populations of
competing self-replicating computer programs \cite{Adam95a}.
In addition to the increasing attention paid to this type of epochal
evolution in the theoretical biology community
\cite{Font98a,Forst&Reidys&Weber95,Huynen95,Newman&Engelhardt98,Reidys98b,Weber97},
recently there has also been an increased interest by evolutionary
search theorists \cite{Barnett97,Haygood}. More directly, Chen et al.
recently proposed to test our original theoretical predictions in an
experimental realization of a genetic algorithm that exhibits epochal
evolution \cite{Chen99a}.

\subsection{Local Optima versus Neutral Subbasins}

How are we to think of the mechanisms that cause epochal evolutionary
behavior? The evolutionary biologist Wright introduced the notion of
``adaptive landscapes'' to describe the (local) stochastic adaptation
of populations to themselves and to environmental fluctuations and
constraints \cite{Wright82a}. This geographical metaphor has had a
powerful influence on theorizing about natural and artificial
evolutionary processes. The basic picture is that of a
gradient-following dynamics moving over a ``landscape'' determined by
a fitness ``potential''. In this view an evolving population
stochastically crawls along a surface determined, perhaps dynamically,
by the fitness of individuals, moving to peaks and very occasionally
hopping across fitness ``valleys'' to nearby, and hopefully higher
fitness, peaks.

More recently, it has been proposed that the typical fitness functions
of combinatorial optimization and biological evolution can be modeled
as ``rugged landscapes'' \cite{Kauf87a,Mack89a}. These are fitness
functions with wildly fluctuating fitnesses even at the smallest
scales of single-point mutations. Consequently, it is generally
assumed that these ``landscapes'' possess a large number of local
optima. With this picture in mind, the common interpretation of
punctuated equilibria in evolving populations is that of a population
being ``stuck'' at a local peak in the landscape, until a rare mutant
crosses a valley of relatively low fitness to a higher peak; a picture
more or less consistent with Wright's.

At the same time, an increasing appreciation has developed, in
contrast to this rugged landscape view, that there are substantial
degeneracies in the genotype-to-phenotype and the phenotype-to-fitness
mappings. The history of this idea goes back to Kimura
\cite{Kimurabook}, who argued that on the genotypic level, most
genetic variation occurring in evolution is adaptively {\em neutral}
with respect to the phenotype. Even today, the crucial role played by
neutrality continues to find important applications in molecular
evolution, for example; see Ref. \cite{Font93a}. During neutral
evolution, when degeneracies in the genotype-phenotype map are
operating, different genotypes in a population fall into a
relatively small number of distinct fitness classes of genotypes with
approximately equal fitness. Due to the high dimensionality of
genotype spaces, sets of genotypes with approximately equal fitness
tend to form components in genotype space that are connected by paths
made of single-mutation steps.

Additionally, due to intrinsic or even exogenous effects (e.g.,
environmental), there simply may not exist a deterministic ``fitness''
value for each genotype. In this case, fluctuations can induce
variation in fitness such that genotypes with similar average fitness
values are not distinct at the level of selection.  Thus,
genotype-to-fitness degeneracies can, to a certain extent, be induced
by noise in the fitness evaluation of individuals.

\begin{figure}[htbp]
\centerline{\epsfig{file=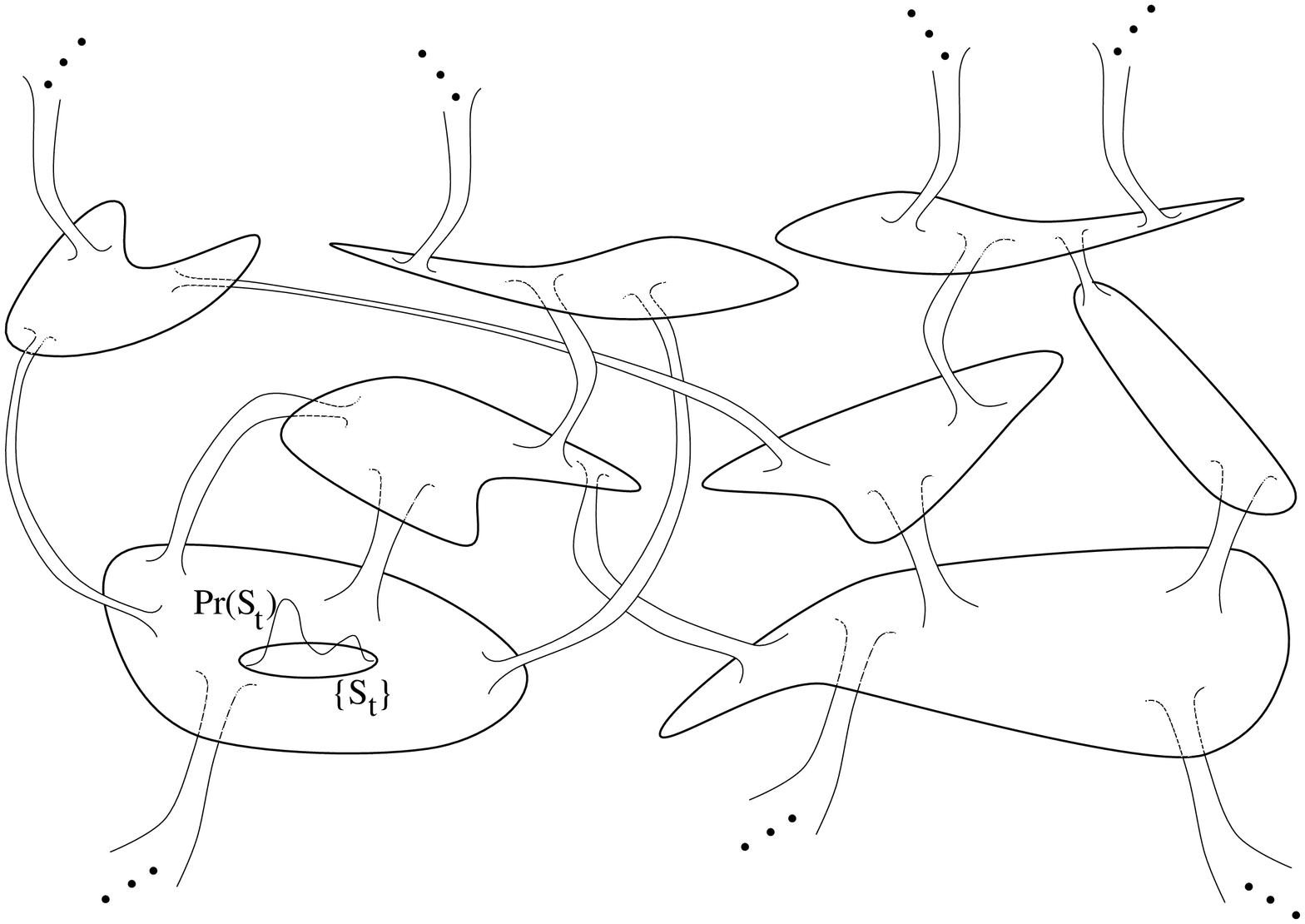,height=3.25in}}
\caption{Subbasin and portal architecture underlying epochal
  evolutionary dynamics. A population---a collection of individuals
  $\{ {\rm S}_t \}$ with distribution $ {\rm Pr} ( {\rm S}_t )$---diffuses
  in the subbasins (large sets) until a portal (tube) to a
  higher-fitness subbasin is found.
  }
\label{SubbasinPortals}
\end{figure}

When these biological facts are taken into account we end up with an
alternative view to both Wright's ``adaptive landscapes'' and the more
recent ``rugged landscapes''. That is, the genotype space decomposes
into a set of neutral networks, or {\em subbasins} of approximately
isofitness genotypes, that are entangled with each other in a
complicated fashion; see Fig. \ref{SubbasinPortals}.
As illustrated in Fig. \ref{SubbasinPortals}, the space of genotypes
is broken into strongly and weakly connected sets with respect to the
genetic operators. Equal-fitness genotypes form strongly connected
neutral subbasins. Moreover, since subbasins of high fitness are
generally much smaller than subbasins of low fitness, a subbasin tends
to be only weakly connected to subbasins of higher fitness.

Since the different genotypes within a neutral subbasin are not
distinguished by selection, neutral evolution---consisting of the
random sampling and genetic variation of individuals---dominates. This
leads to a rather different interpretation of the processes underlying
punctuated equilibria. Instead of the population being pinned at a
local optimum in genotype space as suggested by the ``landscape''
models, the population drifts randomly through neutral subbasins of
isofitness genotypes. A balance between selection and deleterious
mutations leads to a (meta-) stable distribution of fitness (or of
phenotypes), while the population is searching through these spaces of
neutral genotypic variants. Thus, there is no genotypic stasis during
epochs. As was first pointed out in the context of molecular evolution
in Ref. \cite{Huynen&Stadler&Fontana}, through neutral mutations, the
best individuals in the population diffuse over the neutral network of
isofitness genotypes until one of them discovers a connection to a
neutral network of even higher fitness. The fraction of individuals
on this network then grows rapidly, reaching a new equilibrium between
selection and deleterious mutations, after which the new subset of
most-fit individuals diffuses again over the newly discovered neutral
network.

Note that in epochal dynamics there is a natural separation of time
scales. During an epoch selection acts to establish an equilibrium in
the proportions of individuals in the different neutral subspaces, but
it does not induce {\em adaptations} in the population. Adaptation
occurs only in a short burst during an innovation, after which
equilibrium on the level of fitness is re-established in the population.
On a time scale much faster than that between innovations, members of
the population diffuse through subbasins of isofitness genotypes until
a (typically rare) higher-fitness genotype is discovered. Long periods
of stasis occur because the population has to search most of the
neutral subspace before a portal to a higher fitness subspace is
discovered. 

In this way, we shift our view away from the geographic metaphor of
evolutionary adaptation ``crawling'' along a ``landscape'' to the view
of a diffusion process constrained by the subbasin-portal architecture
induced by degeneracies in the genotype-to-phenotype and
phenotype-to-fitness mappings. Moreover, our approach is not simply a
shift towards an architectural view, but it also focuses on the {\em
  dynamics} of populations as they move through the subbasins to find
portals to higher fitness.

\subsection{Epochal Evolution---An Example}
\label{example_ep_dyn}

In our analysis \cite{Nimw97a,Nimw97b}, we view the subbasin-portal
mechanism sketched above as the main source of epochal behavior in
evolutionary dynamics. We will now discuss a simple example of epochal
evolution that illustrates more specifically the mechanisms involved
and allows us to introduce several concepts used in our analysis. 

\begin{figure}[tbp]
\centerline{\epsfig{file=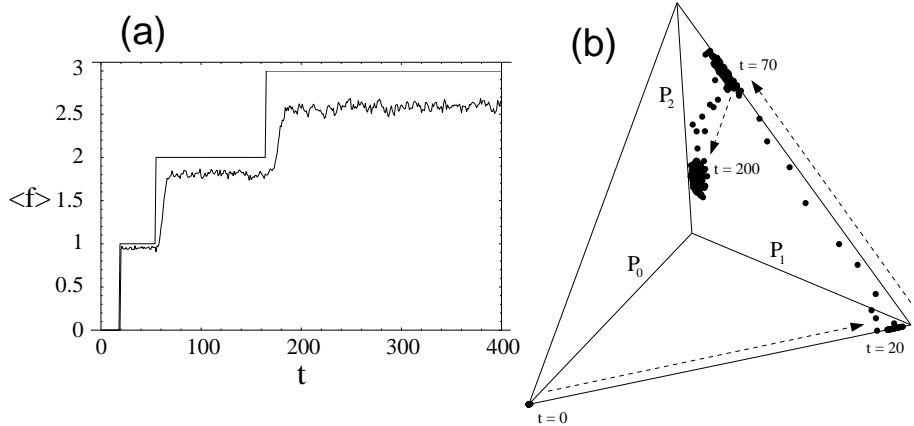,width=4.75in}}
\caption{Dynamics of (a) the average fitness (lower curve)
  and best fitness (upper curve) and
  (b) the fitness distribution for a population evolving under a
  Royal Road fitness function. The fitness function has $N=3$
  constellations of $K=10$ bits each. The population size is $M=250$
  and the mutation rate $\mu = 0.005$. In (b) the location
  of the fitness distribution at each generation is shown
  by a dot. The dashed lines there indicate the direction in which
  the fitness distribution moves from metastable to metastable state
  through the population's fitness-distribution state space (a simplex).
  The times at which the different metastable states are first reached
  are indicated as well.}
\label{FlowandFitness}
\end{figure}

Figure \ref{FlowandFitness} shows the fitness dynamics of an evolving
population on a sample fitness function that exhibits large
degeneracies in the genotype-fitness mapping. This fitness function is
an example of the class of Royal Road fitness functions explained in
Sec.  \ref{fitness_functions} below. The genotype space consists of
all bit-strings of length $30$ and contains neutral subbasins of
fitnesses $0$, $1$, $2$, and $3$. There is only one genotype with
fitness $3$, $3069$ genotypes have fitness $2$, $3.14 \times 10^6$
have fitness $1$, and all others have fitness $0$. The evolving
population consists of $250$ individuals that at each generation are
selected in proportion to their fitness and then mutated with
probability $0.005$ per bit. Figure \ref{FlowandFitness}(a) shows the average
fitness $\langle f \rangle$ in the population (lower curve) and the
best fitness in the population (upper curve) as a function of
generation $t$.

At time $t=0$ the population starts out with $250$ random genotypes.
As can be seen from Fig. \ref{FlowandFitness}(a), during the first few
generations all individuals are located in the largest subbasin with
fitness $0$, since both average and best fitness are $0$. The population
randomly diffuses through this subbasin until, around generation $20$,
a ``portal'' is discovered into the subbasin with fitness $1$. The
population is quickly taken over by genotypes of fitness $1$, until a
balance is established between selection and mutation: selection
expanding and deleterious mutations (from fitness $1$ to $0$) decreasing
the number of individuals with fitness $1$. The individuals with fitness
$1$ continue to diffuse through the subbasin with fitness $1$, until a
portal is discovered connecting to the subbasin with fitness $2$. This
happens around generation $t=60$ and by $t=70$ a new selection-mutation
equilibrium is established. Individuals with fitness $2$ continue
diffusing through their subbasin until the globally optimal genotype
with fitness $3$ is discovered some time around generation $t=170$.
Descendants of this genotype then spread through the population until
around $t=200$, when a final equilibrium is reached.

The same dynamics is plotted in Fig. \ref{FlowandFitness}(b), but from
the point of view of the population's fitness distribution $\vec{P} =
(P_0, P_1, P_2, P_3)$. In the figure the $P_0$ axis indicates the
proportion of fitness $0$ genotypes in the population, $P_1$ the
proportion of fitness $1$ genotypes, and $P_2$ the proportion of
fitness $2$ genotypes. Of course, since $\vec{P}$ is a distribution,
$P_3 = 1-P_0-P_1-P_2$. Due to this, the space of possible fitness
distributions forms a three-dimensional {\em simplex}. We see that
initially $P_0 =1$ and the population is located in the lower-left
corner of the simplex. Later, between $t=20$ and $t=60$, the
population is located at a metastable fixed point on the line $P_0+P_1
=1$ and is dominated by fitness-$1$ genotypes ($P_1 \gg P_0$). Some
time around generation $t=60$ a genotype with fitness $2$ is
discovered and the population moves into the plane $P_0+P_1+P_2 =
1$---the front plane of the simplex. From generation $t=70$ until
generation $t=170$ the population fluctuates around a metastable fixed
point in this plane.  Finally, a genotype of fitness $3$ is discovered
and the population moves to the asymptotically stable fixed point in
the interior of the simplex. It reaches this fixed point around
$t=200$ and remains there fluctuating around it for the rest of the
evolution.

This example illustrates the general qualitative dynamics of epochal
evolution. It is important to note that the architecture of
neutral subbasins and portals is such that a higher-fitness subbasin
is always reachable from the current best-fitness subbasin.
Metastability is a result of the fact that the connections (portals)
to higher-fitness subbasins are very rare. These portals are generally
only discovered after the population has diffused through most of the
subbasin. Additionally, at each innovation, the fitness distribution
expands into a new dimension of the simplex. Initially, when all
members have fitness $0$, the population is restricted to a point.
After the first innovation it moves on a one-dimensional line, after
the second it moves within a two-dimensional plane, and finally it moves
into the interior of the full three-dimensional simplex. One sees that,
when summarizing the population with fitness distributions, the number
of components needed to describe the population grows dynamically each
time a higher-fitness subbasin is discovered. We will return to this
observation when we describe the connection of our analytical approach
to the theory of statistical mechanics.

% ******************* Landscape Architecture *********************

\section{The Terraced Labyrinth Fitness Functions}
\label{fitness_functions}

As just outlined, the intuitive view of phenotypically constrained,
genotype-space architectures---as a relatively small number of weakly
interconnected neutral subbasins---is the one we have adopted in our
analyses. We will now define a broad class of fitness functions that
captures these characteristics. The principal motivation for this is
to illustrate the generality of our existing results via a wider range
of fitness functions than previously analyzed.

We represent genotypes in the population as bit-strings of a fixed
length $L$. For any genotype there is a certain subset of its bits that
are {\em fitness constrained}. Mutations in any of the constrained bits
lowers an individual's fitness. All the other bits are considered
{\em free} bits, in the sense that they may be changed without affecting
fitness. Of all possible configurations of free bits, there is a small
subset of {\em portal configurations} that lead to an increased fitness.
A portal consists of a subset of free bits, called a
{\em constellation}, that is set to a particular ``correct''
configuration. A constellation may have more than one ``correct''
configuration. When a constellation is set to a portal configuration,
the fitness is increased, and the constellation's bits become
constrained bits. That is, via a portal free bits of an incorrectly set
constellation become the constrained bits of a correctly set
constellation.

The general structure of the fitness functions we have in mind is that
fitness is conferred on individuals by having a number of
constellations set to their portal configurations. Mutations in the
constrained bits of the correct constellations lower fitness; while
setting an additional constellation to its portal configuration
increases fitness. A fitness function is specified by choosing sets of
constellations, portal configurations, and assigning the fitness that
each constellation confers on a genotype when set to one of its portal
configurations.

\subsection{A Simple Example}

\begin{figure}[htbp]
\centerline{\epsfig{file=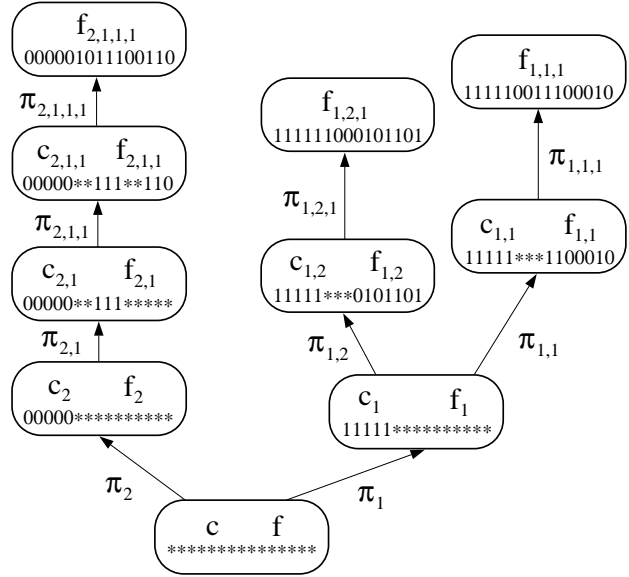,height=3.0in}}
\caption{Tree representation of a Terraced Labyrinth fitness function.
  The nodes of the tree represent subbasins of genotypes with equal
  fitness. They are represented by strings that have $*$'s for the
  free bits. The fitness $f$ of the genotypes in the subbasins is
  indicated as well. The constellation $c$ inside each node
  indicates the subset of bits that needs to be set correctly in order
  to move {\em up} a level in the tree to a higher-fitness subbasin.
  The portal configurations $\pi$ that connect subbasins to
  higher-fitness subbasins are shown as branches.}
\label{FitnessTree}
\end{figure}

Let's illustrate our class of fitness functions by a simple example
that uses bit-strings of length $L=15$. The example is illustrated
in Fig. \ref{FitnessTree}. Initially, when no constellation is set
correctly the strings have fitness $f$. The first constellation,
denoted $c$, consists of the bits $1$ through $5$. This constellation
can be set to two different portal configurations: either
$\pi_1 = 11111$ or $\pi_2 = 00000$. When $c = \pi_1$ or $c = \pi_2$ the
genotypes obtain fitnesses $f_1$ and $f_2$, respectively. Once
constellation $c = \pi_1$, say, there is a constellation $c_1$,
consisting of bits $9$ through $15$, that can be set correctly to
portal configuration $\pi_{1,1} = 1100010$; in which case the genotype
obtains fitness $f_{1,1}$. The constellation $c_1$ might also be set to
configuration $\pi_{1,2} = 0101101$, leading to a fitness of $f_{1,2}$.
Finally, once constellation $c_1 = \pi_{1,1}$, there is a final
configuration $c_{1,1}$, consisting of bits $6$ through $8$, that can
be set correctly. With $c = \pi_1$ and $c_1 = \pi_{1,1}$ configuration
$c_{1,1}$ needs to be set to configuration $\pi_{1,1,1} = 001$ in order
to reach fitness $f_{1,1,1}$. If instead $c_1 = \pi_{1,2}$, the final
constellation $c_{1,2}$ needs to be set to portal $\pi_{1,2,1} = 100$,
giving fitness $f_{1,2,1}$.

Alternatively, if constellation $c = \pi_2$, the next constellation
$c_2$ consists of bits $8$ through $10$, which have portal
configuration $\pi_{2,1} = 111$. Setting $c_2$ to $\pi_{2,1}$ leads to
fitness $f_{2,1}$. Once $c_2$ is set correctly, there is a
constellation $c_{2,1}$ consisting of bits $13$ through $15$, which
has portal configuration $\pi_{2,1,1} = 110$ and fitness $f_{2,1,1}$.
Finally, there is the constellation $c_{2,1,1}$ consisting of bits
$6$, $7$, $11$, and $12$. The portal configuration for this
constellation is $\pi_{2,1,1,1} = 1000$, leading to fitness
$f_{2,1,1,1}$.

%\begin{table}[htbp]
%\centering
%\begin{tabular}{|c|c|c|} \hline
% string type  & correct constellations & fitness \\ \hline
%\texttt{***************} & $\emptyset$ & $f_0$ \\ \hline
%\texttt{11111**********} & $c_0 = \pi_1$ & $f_1$ \\
%\texttt{11111***1100010} & $c_0 = \pi_1 \wedge c_1 = \pi_{1,1}$ &
%$f_{1,1}$ \\
%\texttt{111110011100010} & $c_0 = \pi_1 \wedge c_1 = \pi_{1,1} \wedge
%c_{1,1} = \pi_{1,1,1}$ & $f_{1,1,1}$ \\ \hline
%\texttt{11111***0101101} & $c_0 =\pi_1 \wedge c_1 = \pi_{1,2}$ & $f_{1,2}$ \\
%\texttt{111111001101111} & $c_0 =\pi_1 \wedge c_1 = \pi_{1,2} \wedge
%c_{1,2} = \pi_{1,2,1}$ & $f_{1,2,1}$ \\ \hline
%\texttt{00000**********} & $c_0 = \pi_2$ & $f_2$ \\
%\texttt{00000**111*****} & $c_0 = \pi_2 \wedge c_2 = \pi_{2,1}$ & $f_{2,1}$ \\
%\texttt{00000**111**110} & $c_0 = \pi_2 \wedge c_2 = \pi_{2,1} \wedge
%c_{2,1} = \pi_{2,1,1}$ & $f_{2,1,1}$ \\
%\texttt{000001011100110} & $c_0 = \pi_2 \wedge c_2 = \pi_{2,1} \wedge
%c_{2,1} = \pi_{2,1,1} \wedge c_{2,1,1} = \pi_{2,1,1,1}$ &
%$f_{2,1,1,1}$ \\ \hline
%\end{tabular}
%\caption{A simple example of a Terraced Labyrinth fitness function. The
%  stars indicate dynamic (free) bits.}
%\label{FitnessTree}
%\end{table}

Generally, the hierarchical ordering of constellations and their
connections via portals can be most easily represented as a tree; as in
Fig. \ref{FitnessTree}. Each tree node represents a subbasin of
equal-fitness genotypes. The tree branches represent the portals that
connect a lower-fitness subbasin to a higher-fitness subbasin. The
fitness and structure of genotypes within a subbasin are also
shown at each node. Stars (*) indicate the free bits within a
subbasin. The constellations at each node indicate which subset of
bits needs to be set to a portal configuration in order to proceed
further up the tree. Thus, setting a constellation to a portal
configuration leads one level up the tree, while mutating one or more
of the constrained bits leads down the tree. In fact, a single
point-mutation might lead all the way back to the root node.

We assume that setting a new constellation correctly leads to an
increase in fitness. That is, $f_1$ and $f_2$ are larger than $f$,
$f_{1,1}$ is larger than $f_1$, and so on. For simplicity in this
example, we chose the constellation bits contiguously, except for
$c_{2,1,1}$. Since our genetic algorithm, introduced shortly,
does not employ crossover, the population dynamics remains the
same under arbitrary permutations of the bits in the genome. Note
further that we chose the portal configurations rather arbitrarily. In
cases where a constellation has only a single portal, this
configuration can be chosen arbitrarily without effecting the
dynamics. When a constellation has more than one portal, the
evolutionary dynamics can be affected by the Hamming distances between
the different portal configurations. A key assumption is that portal
configurations such as $\pi_1$ and $\pi_2$ are mutually exclusive.
Once evolution follows a certain branch up
the tree, it is very unlikely to revert later on. We discuss in Sec.
\ref{Accidents} how different evolutionary paths through the tree
formalize such notions as {\em historical accident} and {\em structural
phenotypic constraints}.

Finally, in this setting the genotype-to-phenotype map is nonexistent,
since fitness is evaluated directly on the genotypes, without an
intervening developmental process.

\subsection{Definitions}

We will now generalize this example by way of defining the class of
{\em Terraced Labyrinth} fitness functions. As we saw in the example,
constellations and portals form a hierarchy that can be most easily
represented as a tree. Thus, we define Terraced Labyrinth fitness
functions using trees, similar to the one illustrated in Fig.
\ref{FitnessTree}, as follows.
\begin{enumerate}
\item{The {\em genotypes} are bit strings
${\bf s} = s_0 s_1 s_2 \cdots s_{L-1}$ of length
$L$ with bits $s_i \in {\cal A} \equiv \{0,1\}$.}
\item{The hierarchy of subbasins, constellations, and portals form a
    {\em tree}, consisting of nodes $\{ \vec{\imath} \}$ and
        branches $\{\pi_{\vec{\imath}}\}$.}
\begin{enumerate}
        \item{Tree {\em nodes} $\vec{\imath}$ are specified by a set of
            indices: $\vec{\imath} = \{ i_1,i_2,i_3,\ldots,i_n \}$. The
                number $n$ of indices denotes $\vec{\imath}$'s tree level.
            A particular setting of the indices labels the path from the
                root to $\vec{\imath}$. That is, one reaches $\vec{\imath}$
                by taking branch $i_1$ at the root, branch $i_2$ at node $i_1$,
                and so on. The tree nodes represent both subbasins of genotypes
                with equal fitness and constellations of bits that, when set
                correctly, lead out of one subbasin to the next higher-fitness
                subbasin.}
        \item{Tree {\em branches} represent portal configurations that
                connect the subbasins of equal-fitness genotypes to
                each other. Branch $\pi_{\vec{\imath}}$ points to node
                $\vec{\imath}$.}
\end{enumerate}
\item{A {\em constellation} is a subset of ${\bf s}$'s bits.
        Constellation $c_{\vec{\imath}}$ is located at node $\vec{\imath}$
        and corresponds to the subset of bits that must be set to a portal
    configuration in order to move from subbasin $B_{\vec{\imath}}$ to a
    higher fitness subbasin. The number of bits in a constellation
    $c_{\vec{\imath}}$ is denoted $K_{\vec{\imath}}$.}
\item{A {\em portal} $\pi_{\vec{\imath},j}$ is one particular
    configuration of the $K_{\vec{\imath}}$ bits in
    constellation $c_{\vec{\imath}}$ out of the
    $2^{K_{\vec{\imath}}}$ possible configurations. The indices
    $\vec{\imath}$ of a portal $\pi_{\vec{\imath},j}$ indicate the
        node to which it points.}
\item{The {\em subbasin} $B_{i_1,i_2,\ldots,i_n}$ is the set of
        genotypes that have constellations $c$ through
        $c_{i_1,\ldots,i_{n-1}}$ set to portals $\pi_{i_1}$ through
    $\pi_{i_1,\ldots,i_n}$, respectively, but do {\em not} have
        constellation $c_{i_1,\ldots,i_n}$ set to any of its portal
        configurations.}
\item{All genotypes in the subbasin $B_{\vec{\imath}}$ have a
        fitness $f_{\vec{\imath}}$.}
    \item{A leaf-node $\vec{\imath}$ in the tree represents a set of
        equal-fitness genotypes that form a local optimum of the
        fitness function. The fitness of these genotypes is
        $f_{\vec{\imath}}$.}
\end{enumerate}
The trees that define the hierarchy of constellations, subbasins, and
portals are not entirely arbitrary. They have the following constraints.
\begin{enumerate}
\item{The number of branches leaving node $\vec{\imath}$ is at most
    $2^{K_{\vec{\imath}}}$.}
\item{A constellation is {\em disjoint} from
    the root constellation $c$ and all other constellations that
    connect it to the root. That is, the set $c_{i_1,i_2,\ldots,i_n}$ is
    disjoint from the sets $c$, $c_{i_1}$, $c_{i_1,i_2}$, and so on.}
\end{enumerate}
This class of Terraced Labyrinth fitness functions incorporates and
extends the previously studied {\em Royal Road} fitness functions of
Refs. \cite{Nimw97a} and \cite{Nimw97b} and the {\em Royal Staircase}
fitness functions of Ref. \cite{Nimw98b}. In those fitness functions,
all constellations had the same number of defining bits $K$, and there
was only a single portal configuration $\pi = 1^K$ for each
constellation.  A Royal Staircase fitness function corresponds to a
Terraced Labyrinth fitness function whose tree is a simple linear chain.
Additionally, in the Royal Road fitness functions, constellations were
allowed to be set in any arbitrary order.

%\begin{figure}[htbp]
%\centerline{\epsfig{file=DimensionalSubbasinPortals.epsf,height=3.25in}}
%\caption{The hierarchy of nested subbasins and portals for the Royal
%  Solenoid fitness functions with two portals per subbasin. Here the
%  subbasin-to-subbasin volume decrease is constant; that is,
%  $K_n = K = L/N$. The (high-fitness) attractors are denoted $\Lambda_0$
%  through $\Lambda_7$ and the subbasins of attractors $\Lambda_i$
%  through $\Lambda_j$ are denoted by ${\bf B}_{i-j}$.
%  }
%\label{DimensionalSubbasinPortals}
%\end{figure}

The architectural approach we have taken here should be contrasted
with the use of randomized fitness functions that have been modified
to have neutral networks. These include the NKp landscapes of Ref.
\cite{Barnett97} and the discretized NK fitness functions of Ref.
\cite{Newman&Engelhardt98}. The popularity of random fitness
functions seems motivated by the idea that something as complicated as
a biological genotype-phenotype mapping can only be statistically
described using a randomized structure. Although this seems sensible in
general, the results tend to be strongly dependent on the specific
randomization procedure that is chosen; the results might be
biologically misleading. For instance, NK models create random epistatic
interactions between bits, mimicking spin-glass models in physics. In
the context of spin glasses this procedure is conceptually justified
by the idea that the interactions between the spins were randomly
frozen in when the magnetic material formed. However, in the context
of genotype-phenotype mappings, the interactions between different
genes are themselves the result of evolution. This can lead to very
different kinds of ``random'' interactions, as shown in Ref.
\cite{Alte95a}.

At a minimum, though, the most striking difference
between our choice of fitness function class and randomized fitness
functions, is that the population dynamics of the randomized classes
is very difficult, if not impossible, to analyze at present. In
contrast, the population dynamics of the class of fitness functions
just introduced can be analyzed in some detail.  Moreover, for
biological systems it could very well be that structured fitness
functions, like the Terraced Labyrinth class, may contain all of the
generality required to cover the phenomena claimed to be addressed by
the randomized classes.  Several limitations and generalizations of
the Terraced Labyrinth fitness functions are discussed in Sec.
\ref{extensions_generalizations}.

\section{A Simple Genetic Algorithm}

For our analysis of epochal evolutionary dynamics we chose a simplified
form of a genetic algorithm (GA) that does not include crossover and
that uses fitness-proportionate selection. A population of $M$
individuals, each specified by a genotype of length $L$ bits
reproduces in discrete non-overlapping generations. Each generation,
$M$ individuals are selected (with replacement) from the population in
proportion to their genotype's fitness. Each selected individual is
placed into the population at the next generation after mutating each
genotype bit with probability $\mu$.

This GA effectively has two parameters: the mutation rate $\mu$ and the
population size $M$. A given evolutionary optimization problem is
specified, of course, by the fitness function parameters as given by the
constellations, portals, and their fitness values. Stated most
prosaically, then, our central goal is to analyze the population
dynamics, as a function of $\mu$ and $M$, for any given fitness
function in the Terraced Labyrinth class. Here we
review the essential aspects of the population dynamics analysis.

\section{Statistical Dynamics of Evolutionary Search}

Refs. \cite{Nimw97a} and \cite{Nimw97b} developed an approach,
which we called {\em statistical dynamics}, to analyze the behavioral
regimes of a GA searching fitness functions that lead to epochal
dynamics. Here we can only briefly review the mathematical details of
this approach to evolutionary dynamics, emphasizing
the motivations and the main ideas and tools from statistical mechanics
and dynamical systems theory. The reader is referred to Ref.
\cite{Nimw97b} for an extensive and mathematically detailed exposition.
There, the reader will also find a review of the connections and
similarities of our work with the alternative methodologies for GA
theory developed by Vose and collaborators
\cite{Nix&Vose91,Vose93,Vose&Liepins91}, by Pr\"ugel-Bennett, Rattray,
and Shapiro \cite{PrugelBennett97,PrugelBennett&Shapiro94,Rattray&Shapiro96},
in the theory of molecular evolution
\cite{Eigen71,Eigen&McCasKill&Schuster89}, and in mathematical
population genetics \cite{Hartl&Clark}.

\subsection{Statistical Mechanics}

Our approach builds on ideas from statistical mechanics
\cite{Binn92a,Reic80a,Yeom92a} and adapts its equilibrium formulation
to apply to the piecewise steady-state dynamics of epochal evolution.
The microscopic state of systems that are typically studied in
statistical mechanics---such as, a box of gas molecules---is described
in terms of the positions and momenta of all particles. What is of
physical interest, however, are observable (and reproducible)
quantities, such as, the gas's pressure $P$, temperature $T$, and volume $V$.
The goal is to predict the relationships among these macroscopic
variables, starting from knowledge of the equations of motion
governing the particles and the space of the entire system's possible
microscopic states. A given setting of macroscopic variables---e.g. a
fixed $P$, $V$, and $T$---is often referred to as a {\em macrostate};
whereas a snapshot of the positions and momenta of all particles
is called a {\em microstate}.

There are two kinds of assumptions that allow one to connect the
microscopic description (collection of microstates and equations of
motion) to observed macroscopic behavior. The first is the assumption
of {\em maximum entropy} which states that all microscopic variables,
unconstrained by a given macrostate, are as random as possible.

The second is the assumption of {\em self-averaging}. In the {\em
  thermodynamic limit} of an infinite number of particles,
self-averaging says that the macroscopic variables are expressible
only in terms of themselves. In other words, the macroscopic
description does not require knowledge of detailed statistics of the
microscopic variables. For example, at equilibrium the macroscopic
variables of an {\em ideal gas} of noninteracting particles are
related by the {\em equation of state}, $P V = k N T$, where $k$ is a
physical constant, and $N$ is the total number of particles in the
box. Knowing, for instance, the frequency with which molecules come
within $100$ nanometers of each other does not improve this macroscopic
description.

Varying an experimental control parameter of a thermodynamic system
can lead to a sudden change in its structure and in its macroscopic
properties. This occurs, for example, as one lowers the temperature of
liquid water below the freezing point. The liquid macrostate undergoes
a {\em phase transition} and the water turns to solid ice. The
macrostates ({\em phases}) on either side of the transition are
distinguished by different sets of macroscopic variables. That is, the
set of macrovariables that is needed to describe ice is not the same
as the set of macrovariables that is needed to describe water. The
difference between liquid water and solid ice is captured by a sudden
reduction in the freedom of water molecules to move. While the water
molecules move equally in all directions, the frozen molecules
in the ice-crystal possess a relatively definite spatial location.
Passing through a phase transition can be thought of as creating, or
destroying, macroscopic variables and making or {\em breaking} the
symmetries associated with them. In the liquid to solid transition,
the rotational symmetry of the liquid phase is broken by the onset of
the rigid lattice symmetry of the solid phase. As another example, in
the Curie transition of a ferromagnet, the {\em magnetization} is the
new macroscopic variable that is created with the onset of
magnetic-spin alignment as the temperature is lowered. 

\subsection{Evolutionary Statistical Mechanics}

The statistical mechanical description can also be applied to
evolutionary processes. From a microscopic point of view, the exact
state of an evolving population is only fully described when a list
$\cal{S}$ of all genotypes with their frequencies of occurrence in the
population is given. On the microscopic level, the evolutionary
dynamics is implemented as a {\em Markov chain} with the conditional
transition probabilities ${\rm Pr}(\cal{S'} | \cal{S})$ that the
population at the next generation will be the ``microscopic''
collection $\cal{S'}$; see Refs.  \cite{Ewensbook} and
\cite{Nix&Vose91} for the microscopic formulation in the context of
mathematical population genetics and genetic algorithms, respectively.
For any reasonable genetic representation, however, there is an
enormous number of these microscopic states $\cal{S}$ and so too of
their transition probabilities. The large number of parameters, ${\cal
  O}(2^L!)$, makes it almost impossible to quantitatively study the
dynamics at this microscopic level.

More practically, a full description of the dynamics on the
level of microscopic states $\cal{S}$ is neither useful nor typically
of interest. One is much more likely to be concerned with relatively
coarse statistics of the dynamics, such as the evolution of the best
and average fitness in the population or the waiting times for evolution
to produce a genotype of a certain quality. The result is that
quantitative mathematical analysis faces the task of finding a
macroscopic description of the microscopic evolutionary dynamics
that is simple enough to be tractable numerically or analytically and
that, moreover, facilitates predicting the quantities of interest to
an experimentalist.

With these issues in mind, we specify the macrostate of the population
at each time $t$ by some relatively small set of macroscopic variables
$\{ {\cal X} (t) \}$. Since this set of variables intentionally
ignores vast amounts of detail in the microscopic variables $\{ {\rm
  x}(t)\}$, it is generally impossible to exactly describe the
evolutionary dynamics in terms of these macroscopic variables. To
achieve the benefits of a coarser description, we assume that the
population has equal probabilities to be in any of the microscopic
states consistent with a given macroscopic state. That is, we assume
{\em maximum entropy} over all microstates $\{ {\rm x}(t)\}$ that are
consistent with the specific macrostate $\{ {\cal X} (t) \}$.

Additionally, in the limit of infinite-population size, we assume that
the resulting equations of motion for the macroscopic variables become
closed. That is, for infinite populations, we assume that we can
predict the state of the macroscopic variables at the next generation,
given the present state of {\em only} the macroscopic variables. This
infinite population limit is analogous to the thermodynamic limit in
statistical mechanics. The corresponding assumption is analogous to
{\em self-averaging} of the macroscopic evolutionary dynamics in this
limit.

We use the knowledge of the microscopic dynamics together with the
maximum entropy assumption to predict the next macrostate $\{ {\cal X}
(t+1) \}$ from the current one $\{ {\cal X} (t) \}$. Then we re-assume
maximum entropy over the microstates $\{ {\rm x}(t+1)\}$ given the new
macrostate $\{ {\cal X} (t+1) \}$. Since this method allows one to
relax the usual equilibrium constraints and so account for the
dynamical change in macroscopic variables, we refer to this extension
of statistical mechanics as {\em statistical dynamics}. A similar
approach has been developed in some generality for non-equilibrium
statistical mechanics by Streater and, not surprisingly, it goes
under the same name \cite{Stre95a}.

\subsection{Evolutionary Macrostates}

The key, and as yet unspecified step, in developing such a statistical
dynamics framework of evolutionary processes is to find an appropriate
set of macroscopic variables that satisfy the above assumptions of
maximum entropy and self-averaging. In practice, this is difficult.
Ultimately, the suitability of a set of macroscopic variables has to
be verified by comparing theoretical predictions with experimental
measurements. In choosing such a set of macroscopic variables one is
guided by knowledge of the fitness function and the genetic operators.
Although not reduced to a procedure, this choice is not made in the
dark.

First, there might be {\em symmetries} in the microscopic dynamics.
Imagine, for instance, that genotypes can have only two possible
values for fitness, $f_A$ and $f_B$. Assume also that under mutation
all genotypes of type $A$ are equally likely to turn into type-$B$
genotypes and that all genotypes of type $B$ have equal probability to
turn into genotypes of type $A$. In this situation, it is easy to see
that we can take the macroscopic variables to be the relative
proportions of $A$ genotypes and $B$ genotypes in the population. The
reason one can do this is that {\em all} microstates with a certain proportion of $A$
and $B$ types give rise to exactly the same dynamics on the level of
proportions of $A$ and $B$ types. That is, the dynamics is symmetric
under any transformation of the microstates that leaves the proportions
of $A$ and $B$ types unaltered. Neither selection nor mutation
distinguish different genotypes within the sets $A$ and $B$ on the level
of the proportions of $A$'s and $B$'s that they produce in the next
generation. Obviously, one wants to take advantage of such symmetries in
a macroscopic description. However, for realistic cases, such symmetries
are not often abundant. Simply taking them into account, while
important, does not typically reduce the complexity of the description
sufficiently. 

One tends to make more elaborate assumptions in developing a macroscopic
description. Assume that the $A$ and $B$ genotypes are not all equally
likely to turn from type $A$ to $B$ and vice versa, but do so only on
{\em average}. For example, it might be the case that not all $A$ types
behave exactly the same under mutation, but that the dominant subset of
$A$'s that occurs in a population typically behaves like the
{\em average} over the set of all $A$ types. This is a much weaker
symmetry than the exact one mentioned above. Importantly, it still
leads to an accurate description of the dynamics on the level of
$A$ and $B$ types under the maximum entropy assumption.

The Neo-Darwinian formalism of biological evolution suggests a natural
decomposition of the microscopic population dynamics into a part that
is guided by selection and a part that is driven by genetic
diversification. Simply stated, selection is an ordering force induced
by the environment that operates on the level of the phenotypic fitness
in a population. In contrast, genetic diversification is a disordering
and randomizing force that drives a population to an increased diversity
of genotypes. Thus, it seems natural to choose as macrostates the
proportion of genotypes in the different fitness classes (subbasins)
and to assume that, due to random genetic diversification within each
subbasin, genetic variation can be approximated by the maximum entropy
distribution within each subbasin. This intuition is exactly the one we
use in our statistical dynamics analysis of the Terraced Labyrinth
fitness functions. Specifically, we describe the population in terms
of the proportions $P_{\vec{\imath}}$ that are located in each of the
subbasins $B_{\vec{\imath}}$. The maximum entropy assumption entails
that within subbasin $B_{\vec{\imath}}$, individuals are equally likely
to be any of the genotypes in $B_{\vec{\imath}}$. In other words, we
assume that all free bits in a constellation are equally likely to be
in any of their nonportal configurations.

The essence of our statistical dynamics approach is to describe
the population state at any time during a GA run by a relatively small
number of macroscopic variables---variables that (i) in the limit of
infinite populations self-consistently describe the dynamics at their
own level and (ii) can change over time. After obtaining the dynamics
in the limit of infinite populations explicitly, one then uses this
knowledge to solve for the GA's dynamical behaviors with finite
populations.

\section{Evolutionary Dynamical Systems}
\label{EvDynSys}

Up to this point we have described our approach in terms of its
similarities with statistical mechanics. We appealed intuitively to
macroscopic ``dynamics'', which can be derived in terms of the
microscopic equations of motion (of selection and mutation on
genotypes) and the maximum entropy assumption. Now we fill in the
other half of the story, the half that clarifies what ``dynamics'' is
and that draws out the similarities of our approach with dynamical
systems theory.

As we just explained, we approximate the complete finite-population
dynamics in two steps. First, we use the maximum entropy assumption
together with the microscopic equations of motion to construct an
infinite-population ``flow'' that describes the deterministic
(macroscopic) dynamics of the subbasin distribution of an infinite
population. Then, we construct the finite-population dynamics by
accounting for the finite-population sampling at each generation. The
net result is a stochastic nonlinear dynamical system. We now explain
these two steps in more detail.

\subsection{Infinite Populations}
\label{InfinitePopulations}

Consider an infinite population with subbasin distribution $\vec{P}$,
where component $P_{\vec{\imath}} \in[0,1]$ is the proportional of
individuals in the subbasin $B_{\vec{\imath}}$. Note that the number
of components in $\vec{P}$ is equal to the number of nodes in the
constellation tree that describes the Terraced Labyrinth fitness
function. Given this, the question is how selection and mutation,
acting on the distribution $\vec{P}(t)$, create the distribution
$\vec{P}(t+1)$ at the next generation.

The effects of selection are simple, since all genotypes in subbasin
$B_{\vec{\imath}}$ have the same fitness. If $\langle f \rangle$ is
the average fitness in the population, we simply have that after
selection the components are
$P^{select}_{\vec{\imath}} = f_{\vec{\imath}} P_{\vec{\imath}}(t)/\langle f \rangle$.
To calculate
the effects of mutation we have to use our maximum entropy assumption.
The probability that a genotype in subbasin $B_{\vec{\jmath}}$ turns
into a genotype in subbasin $B_{\vec{\imath}}$ is simply given by the
{\em average} probability of a mutation from a genotype in
$B_{\vec{\jmath}}$ to {\em any} genotype in $B_{\vec{\imath}}$.
The average is taken with equal weights over all genotypes in
$B_{\vec{\jmath}}$. Putting the effects of selection and mutation
together, we obtain a generation operator ${\bf G}$ that
specifies the macroscopic evolutionary dynamical system:
\begin{equation}
\vec{P}(t+1) = {\bf G} [ \vec{P}(t) ]~.
\label{MacroDynamics}
\end{equation}
The infinite population dynamics on the level of subbasin
distributions is simply given by iterating the operator ${\bf G}$. 
Following the terminology introduced in molecular evolution theory we
call $\vec{P}(t)$ the {\em phenotypic quasispecies}.

\begin{figure}[htbp]
\centerline{\epsfig{figure=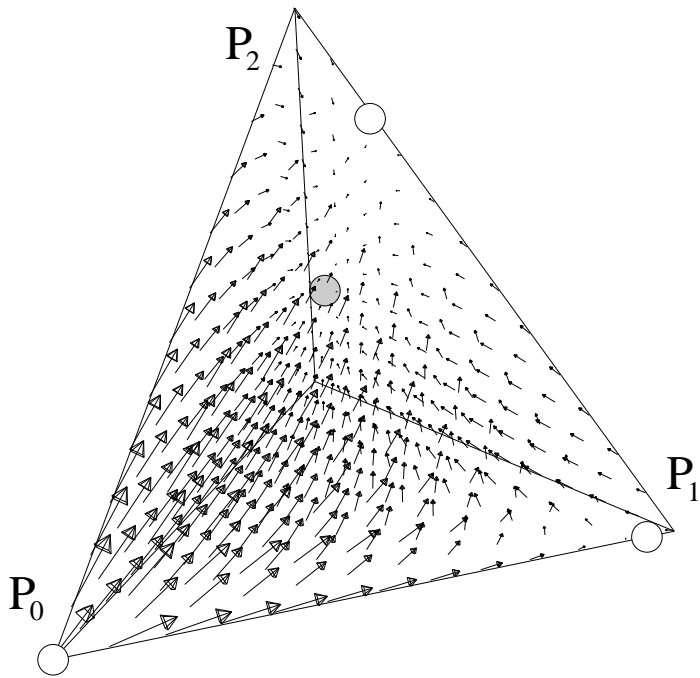,height=3.5in}}
\caption{Fitness distribution flow $\langle d\vec{P} \rangle$ in the
  simplex for the Royal Road fitness function with $N=3$
  constellations with $K=10$ bits each and for the simple GA with
  mutation rate $\mu=0.005$; cf. Fig. \ref{FlowandFitness}. Fixed
  points of the flow are shown as large balls. The grey ball is the
  stable, asymptotic fixed point inside the simplex. The white balls
  indicate the locations of the unstable fixed points that are outside
  the simplex. The latter do not represent valid populations, but
  nonetheless they can affect the dynamics of allowed populations
  within the simplex by slowing down (short arrows) the flow near
  them.}
\label{flow}
\end{figure}

The expected\footnote{It will become clear shortly why we call this
  change an {\em expected} change.} change $\langle d\vec{P} \rangle$
in the fitness distribution over one generation is given by:
\begin{equation}
\langle d\vec{P}\rangle = {\bf G} [ \vec{P} ] -\vec{P}.
\end{equation}
We visualize the {\em flow} induced by the macroscopic equations of
motion by plotting $\langle d\vec{P} \rangle$ at a number of states in
the simplex of populations. This is shown in Fig. \ref{flow}; after Ref.
\cite{Nimw97b}. The fitness function and evolution parameters of Fig.
\ref{flow} are those of Fig. \ref{FlowandFitness}. The temporal behavior
of the system, starting in an initial condition $\vec{P}(t=0)$, is
simply given by following the flow arrows.

For large ($M > 2^L$) populations the dynamics of the subbasin
distribution is simple: $\langle f \rangle$ increases smoothly and
monotonically to an asymptote over a small number of generations. (See
Fig. 3 of Ref. \cite{Nimw97a}.) That is, there are no epochs. The
reason for this is simple: for an infinite population, {\em all}
genotypes, and therefore all subbasins, are represented in the initial
population. Instead of the evolutionary dynamics {\em discovering}
fitter genotypes over time, it essentially only expands the proportion
of globally optimal genotypes already present in the initial
population at $t=0$.

\subsection{Finite Populations}
\label{FinitePopulations}

In spite of the qualitatively different dynamics for infinite and
finite populations, we showed in Ref. \cite{Nimw97b} that the
(infinite population) operator ${\bf G}$ is the essential ingredient
for describing the finite-population dynamics with its epochal
dynamics as well. Beyond the differences in observed behavior, there
are two important mathematical differences between the
infinite-population dynamics and that with finite populations. The
first is that with finite populations the components
$P_{\vec{\imath}}$ cannot take on continuous values between $0$ and
$1$. Since the number of individuals in subbasin $B_{\vec{\imath}}$ is
necessarily an integer, the values of $P_{\vec{\imath}}$ are quantized
in multiples of $1/M$. Thus, the continuous simplex of allowed
infinite-population fitness distributions turns into a regular,
discrete lattice with spacing of $1/M$. Second, due to finite-population
sampling fluctuations, the dynamics of the subbasin
distribution is no longer deterministic, as described by
Eq. (\ref{MacroDynamics}). In general, we can only
determine the conditional probabilities ${\rm Pr}[\vec{Q}|\vec{P}]$
that a given fitness distribution $\vec{P}$ leads to another $\vec{Q}$
in the next generation.

\begin{figure}[htbp]
\centerline{\epsfig{file=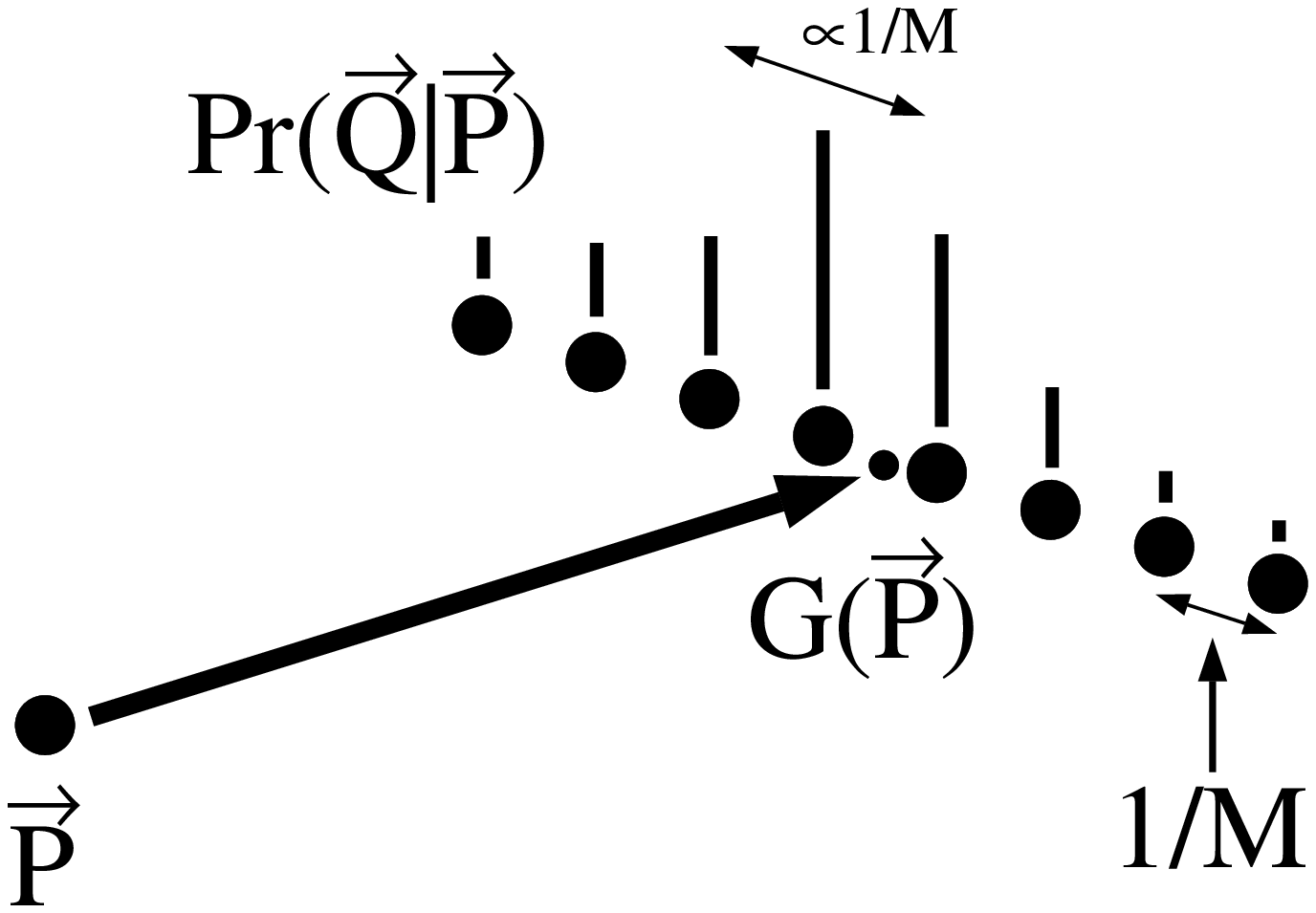,width=3.0in}}
\caption{Illustration of the stochastic dynamics that maps from
  one generation to the next. Starting with finite population
  $\vec{P}$, the arrow indicates the {\em expected} next population
  ${\bf G} [\vec{P}]$. Due to sampling, the probability that the
  actual next population is $\vec{Q}$ is given by a multinomial
  distribution ${\rm Pr}[\vec{Q}|\vec{P}]$, Eq. (\ref{SamplingDist}).
  Note that the underlying state space is a discrete lattice with
  spacing $1/M$.}
\label{SamplingDynamics}
\end{figure}

The net result is that the probabilities ${\rm Pr}[\vec{Q}|\vec{P}]$
are determined by a multinomial distribution with mean ${\bf G} [
\vec{P} ]$:
\begin{equation}
{\rm Pr} [ \vec{Q} | \vec{P} ] = M!
\prod_{\vec{\imath}} \frac{\left({\bf G}_{\vec{\imath}} [ \vec{P} ]
  \right)^{m_{\vec{\imath}}}}{m_{\vec{\imath}}!} ~.
\label{SamplingDist}
\end{equation}
where $Q_{\vec{\imath}} = m_{\vec{\imath}}/M$, with $0 \leq
m_{\vec{\imath}} \leq M$ integers and the product runs over all
subbasins $\vec{\imath}$. (The stochastic effects of
finite-population sampling are illustrated in Fig.
\ref{SamplingDynamics}.) For any finite-population subbasin
distribution $\vec{P}$ the operator ${\bf G}$ gives the evolution's
{\em average} dynamics over one time step, since by Eq.
(\ref{SamplingDist}) the {\em expected} subbasin distribution at the
next time step is ${\bf G} [ \vec{P} ]$.  Note that the components
${\bf G}_{\vec{\imath}} [ \vec{P} ]$ need not be multiples of $1/M$.
Therefore, the actual subbasin distribution $\vec{Q}$ at the next time
step is not ${\bf G} [ \vec{P} ]$, but is instead one of the allowed
lattice points in the finite-population state space consistent with
the distribution ${\rm Pr}[\vec{Q}|\vec{P}]$. Since the variance
around the expected distribution ${\bf G}[\vec{P}]$ is proportional to
$1/M$, $\vec{Q}$ tends to be one of the lattice points close to ${\bf
  G} [ \vec{P} ]$.

Putting both the infinite-population dynamical system and the stochastic
sampling effects induced by finite populations together, we arrive at
the our basic model of evolutionary population dynamics. We can now
begin to draw out some consequences.

\section{Metastability and the Unfolding of Macrostates}
\label{EvUnfold}

Assume that there are no individuals in a certain subbasin
$B_{\vec{\imath}}$ and that the component $\langle d P_{\vec{\imath}}
\rangle$ is much smaller than $1/M$. In that case, the actual change
in component $P_{\vec{\imath}}$ is likely to be $d P_{\vec{\imath}} =
0$ for a long succession of generations. That is, if there are no
individuals in subbasin $B_{\vec{\imath}}$ and the rate of creation of
such individuals is much smaller than $1/M$, then subbasin
$B_{\vec{\imath}}$ is likely to stay empty for a considerable number
of generations. Consequently, there is no movement to increase fitness
to level $f_{\vec{\imath}}$ during this time. More generally, if the
size of the flow $\langle dP_{\vec{\imath}}\rangle$ (and its variance)
in some direction $\vec{\imath}$ is much smaller than the lattice
spacing ($1/M$) of allowed finite populations, we expect the subbasin
distribution to not change in direction $\vec{\imath}$. In Refs.
\cite{Nimw97a} and \cite{Nimw97b} we showed this is the mechanism that
causes epochal dynamics for finite populations.

More formally, an epoch corresponds to the population being restricted
to a region of an $n$-dimensional subsimplex of the macroscopic state
space. Stasis occurs because the flow out of this subspace is much
smaller than the finite-population induced lattice spacing. In
particular, for the Terraced Labyrinth fitness functions, an epoch
corresponds to the time during which the highest fitness individuals
are located in subbasin $B_{i_1,i_2,\ldots,i_n}$. During this time, an
equilibrium subbasin distribution is established in the
population. Its components are nonzero only for subbasins $B$,
$B_{i_1}$, $B_{i_1,i_2}$, through $B_{i_1,\ldots,i_n}$. That
is, they are nonzero for all of the lower
fitness subbasins that connect $B_{\vec{\imath}}$ to the root. Since
the discovery of a portal configuration of constellation
$c_{i_1,\ldots,i_n}$ is rare, the population remains in this
$n$-dimensional subsimplex for a considerable number of generations.
The number of generations it remains in this epoch is, of course,
directly dependent on the number of portals out of the subbasin
$B_{\vec{\imath}}$ and the number of bits $K_{\vec{\imath}}$ in
constellation $c_{\vec{\imath}}$.

Recall the example of epochal behavior of Sec. \ref{example_ep_dyn}
and Fig. \ref{FlowandFitness}. Initially, the population was located
in the zero-dimensional macrostate corresponding to all genotypes
located in the root subbasin. Then the first portal configuration was
discovered and the population moved onto the line of population states
that have some individuals in the root subbasin and some in the basin
$B_{1}$. After this epoch, a genotype in subbasin $B_{1,1}$ was
discovered and the population moved to a steady-state in the plane of
proportions $P$, $P_1$, and $P_{1,1}$. (These were labeled according to
their fitnesses---$P_0$, $P_1$, and $P_2$---in Fig. \ref{FlowandFitness}.)
Finally, the global optimum string in subbasin $B_{1,1,1}$ was
discovered, and the population moved to its final fixed point in the
three-dimensional simplex.

The global evolutionary dynamics can be viewed as an incremental
discovery (an unfolding) of successively more (macroscopic) dimensions of the
subbasin distribution space.  In most realistic settings, it is
typically the case that population sizes $M$ are much smaller than
$2^L$. Initially, then, the population consists only of genotypes in
subbasins of low fitness. Assume, for instance, that genotypes in subbasin
$B_{1,2}$ are the highest fitness ones in the initial population.
Mutation and selection establish an equilibrium phenotypic
quasispecies $\vec{P}^{1,2}$, consisting of nonzero proportions of
genotypes in the subbasin $B$, $B_1$, and $B_{1,2}$, and zero
proportions of genotypes in all other subbasins. Individuals and their
descendants drift through subbasin $B_{1,2}$.  The
subbasin distribution fluctuates around $\vec{P}^{1,2}$ until a portal
configuration $\pi_{1,2,i}$ of the constellation $c_{1,2}$ is
discovered and genotypes of (higher) fitness $f_{1,2,i}$ spread through
the population.  The population then settles into subbasin
distribution $\vec{P}^{1,2,i}$ with average fitness $\langle f
\rangle_{1,2,i}$ until a portal $\pi_{1,2,i,j}$ of constellation
$c_{1,2,i}$ is discovered, and so on, until a local optimum
corresponding to a leaf of the fitness function tree is found. In this
way, the macroscopic dynamics can be seen as stochastically hopping
between the different epoch distributions $\vec{P}^{\vec{\imath}}$ of
subbasins $B_{\vec{\imath}}$ that are connected to each other in the
fitness function tree.

Note that at each stage $\vec{P}^{i_1,\ldots,i_n}$ has only $n+1$
(nonzero) components, each corresponding to a subbasin connecting
$B_{\vec{\imath}}$ to the tree root. All other subbasin components are
zero. The selection-mutation balance maintains a constant proportion
of genotypes with correct configurations in all constellations that
define the epoch. By the maximum entropy assumption, the action of the
generation operator $\bf G$ is symmetric with respect to all remaining
nonportal constellation configurations. That is, $\bf G$'s action is
indifferent to the various proportions of particular incorrect
constellations configurations. The symmetry among constellation
$c_{\vec{\imath}}$'s incorrect configurations is {\em broken
  dynamically} when a (typically, rare) portal configuration is
discovered. This symmetry breaking adds a new macroscopic variable---a
new ``active'' dimension of the phenotype. This symmetry breaking and
stabilization of a new phenotypic dimension is the dynamical analogue of
a phase transition.

As alluded to earlier, much of the attractiveness of the Terraced
Labyrinth class of fitness functions lies in the fact that, to a good
approximation, analytical predictions can be obtained for observable
quantities; such as, average epoch fitness $\langle f
\rangle_{\vec{\imath}}$ and the epoch subbasin distribution
$\vec{P}^{\vec{\imath}}$ in terms of the evolutionary and fitness
function parameters. For instance, assume that the highest fitness
genotypes are in subbasin $B_{i_1,i_2,\ldots,i_n}$ and that the
population resides in the steady-state distribution
$\vec{P}^{i_1,i_2,\ldots,i_n}$. Denote by
\begin{equation}
L_{i_1,i_2,\ldots,i_m} = K + K_{i_1} + K_{i_1,i_2} + \ldots + K_{i_1,i_2,\ldots,i_{m-1}},
\end{equation}
the number of constrained bits in each of the subbasins that have
nonzero proportions during this epoch. (Note that $L=0$ for the root
subbasin).  Then, up to some approximation,\footnote{The approximation
here is that, during an epoch, the {\em back mutations} from lower
fitness subbasins to higher subbasins can be neglected. This
assumption is generally valid for constellation lengths
$K_{\vec{\imath}}$ that are not too small.} the average epoch
fitness is simply given by
\begin{equation}
\langle f \rangle_{\vec{\imath}} = f_{\vec{\imath}} \, (1-\mu)^{L_{\vec{\imath}}}.
\label{EpochFitness}
\end{equation}

One can also derive the subbasin distribution $\vec{P}^{\vec{\imath}}$.
In order to express the results most transparently, we introduce the
fitness-level ratio using Eq. (\ref{EpochFitness}):
\begin{equation}
\alpha_{\vec{\imath}\vec{\jmath}} =
\frac{f_{\vec{\jmath}}}{f_{\vec{\imath}}}
(1-\mu)^{L_{\vec{\jmath}}-L_{\vec{\imath}}}
\end{equation}
Then we have for the highest-fitness component of the subbasin
distribution $P_{\vec{\imath}}$ that
\begin{equation}
P^{\vec{\imath}}_{\vec{\imath}} = \prod_{\vec{m} < 
  \vec{\imath}}
  \frac{1-\alpha_{\vec{\imath}\vec{m}}}{1-\alpha_{\vec{\imath}\vec{m}}
  (1-\mu)^{K_{\vec{m}}}},
\end{equation}
where $\vec{m} < \vec{\imath}$ indicates the set of all nodes lying
along the path between $\vec{\imath}$ and the tree's root, including
the root. For the other components of $P_{\vec{\imath}}$ we have that
\begin{equation}
P^{\vec{\imath}}_{\vec{\jmath}} = \frac{(1-\mu)^{L_{\vec{\jmath}}} \left(1 -
  (1-\mu)^{K_{\vec{\jmath}}} \right)}{1 -
\alpha_{\vec{\imath}\vec{\jmath}} (1-\mu)^{K_{\vec{\jmath}}}}
  \prod_{\vec{m} < \vec{\jmath}} \frac{1-\alpha_{\vec{\imath}\vec{m}}}{1-\alpha_{\vec{\imath}\vec{m}}
  (1-\mu)^{K_{\vec{m}}}}.
\end{equation}
Describing the dynamics in and between epoch distributions
$\vec{P}^{\vec{\imath}}$ using diffusion approximations and then
invoking (dynamical systems) concepts---such as, stable and unstable
manifolds, Jacobian eigenvalues, and their eigenvectors---a number of
additional properties of epochal evolution can be derived analytically
and predicted quantitatively. The reader is referred to Refs.
\cite{Nimw97b} and \cite{Nimw98b} for the detailed analysis of the
distribution of epoch fluctuations, the stability of epochs, and the
average waiting times for portal discovery.

\section{Frozen Accidents, Phenotypic Structural Constraints, and
the Subbasin-Portal Architecture}
\label{Accidents}

The subbasin-portal architecture, whose population dynamics we are
analyzing, suggests a natural explanation for the occurrence and
longevity of {\em frozen accidents}. Generally speaking, frozen
accidents refer to persistent phenotypic characters that are selected
out of a range of possible, structurally distinct alternatives by
specific random events in the evolutionary past. One imagines an
arbitrary event, such as a sampling fluctuation, promoting one or
another phenotype, which then comes to dominate the population and
thereby excludes alternatives that could be equally or even more fit
in the long term. 

Within the class of Terraced Labyrinth fitness functions {\em frozen
accidents} occur via a simple mechanism. In particular, a given
evolutionary path through the fitness-function tree can be regarded as
a sequence of frozen accidents. Since different portals of the same
constellation are mutually exclusive, their subbasins are separated by
a {\em fitness barrier}. Across a wide range of parameter settings, the
crossing of such fitness barriers takes much longer than the discovery
of new portals, via neutral evolution, in the current subbasin. Once
evolution has taken a certain branch up the tree, it is therefore
unlikely, that it will ever return. That is, once a subbasin
$B_{\vec{\imath}}$ is discovered, the further course of evolution is
restricted to the subtree with its root at $\vec{\imath}$. In this way,
the genotypic constellations up to $\vec{\imath}$ become installed in
the population.

The alternative evolutionary paths are not merely a case of genetic
bookkeeping. Different portals of a constellation $c_{\vec{\imath}}$
may be associated with very different {\em phenotypic} innovations.
Once a particular phenotypic innovation has occurred, the phenotype
determines which range of future phenotypic innovations can
occur. This contingency---how evolutionary futures depend on
current phenotypic constraints---goes under the name of
{\em structural phenotypic constraints}. In the Terraced Labyrinth this
phenomenon is reflected in the possibility that fitness-function trees
have very dissimilar subtrees. For instance the subtrees rooted at
nodes $1$ and $2$ in Fig. \ref{FitnessTree} are very dissimilar. This
dissimilarity reflects the fact that evolutionary futures starting from
the phenotype corresponding to node $1$ are very different from those
starting from the phenotype associated with node $2$.

Naturally, the Terraced Labyrinth class of fitness functions does not
indicate which kind of tree structures, reflecting structural
constraints, are appropriate or biologically realistic. This will
ultimately be decided by experiment. The generality of this class of
fitness functions, however, illustrates that qualitative
concepts---such as, frozen accidents and structural phenotypic
constraints---are very easily represented and analyzed within the
statistical dynamics framework.

\section{Concluding Remarks}

\subsection{Summary}

We introduced a generalized subbasin-portal architecture by way of
defining a new class of fitness functions---the Terraced Labyrinth. The
detailed mathematical analysis of the population dynamics that we
introduced previously can be adapted straightforwardly to this
generalized setting. In this way, statistical dynamics was shown to
have a wider applicability and its results on epochal evolution are
seen to have wider ranging consequences than the first analyses in
Refs. \cite{Nimw97a} and \cite{Nimw97b} might have suggested.

We described this more general view of epochal evolution, attempting
to clarify the connections to both statistical mechanics and dynamical
systems theory. The result is a dynamical picture of a succession of
``phase transitions'' in which microscopic symmetries are broken and
new macroscopic dimensions are discovered and then stabilized. These
new macroscopic dimensions then become the substrate and historical
context for further evolution.

\subsection{Extensions and Generalizations}
\label{extensions_generalizations}

There are a number of extensions to more complex evolutionary
processes that should now be possible. Here we mention a few
limitations of the class of fitness functions analyzed and
several generalizations.

First, constellations do not overlap constellations higher in the
tree.  Second, all the subbasins have a similar regular architecture:
there is a set of constrained bits (in the portals) that define the
subbasin and all other bits are free.This is undoubtedly not the case
generally. Different subbasins can have distinct irregular
architectures and different kinds of portals. Moreover, the diffusion
dynamics through distinct subbasins might be different. For instance,
subbasins might also be defined with respect to more complicated
genetic operations---such as, gene duplication, unequal crossovers,
and gene conversion.

Third, all of a subbasin's portals correspond to configurations of a
single constellation. This insures that the topology of the subbasin
hierarchy forms a tree, as opposed to the more general topologies
suggested by Fig. \ref{SubbasinPortals}. Extending the analysis to
more complicated subbasin architectures is formally straightforward,
but becomes considerably more complicated to carry out. For very
complicated architectures, the approximations in our analysis may
have to be reworked.

Fourth, one would like to extend statistical dynamics to {\em
open-ended} models in which (say) the genotype length can grow over
time, allowing the tree to {\em dynamically} grow new branches as
well; perhaps along the lines investigated in Ref. \cite{Alte95a}. One
would hope to see how the evolutionary dynamics adapts as the
mutation-genome length error threshold is approached \cite{Eigen71}.
As long as such open-ended models adhere to the tree topology of the
subbasin-portal hierarchy, it would appear that our analyses could
easily be extended to them.

Finally, the maximum entropy assumption only holds to some degree of
approximation. For instance, whenever a new macrodimension unfolds,
the population is initially concentrated around the portal genotype in
the neutral network; this is a type of {\em founder effect}. The
population then spreads out randomly from there, but the genotypes
never completely decorrelate due to finite-population sampling
fluctuations \cite{Derrida&Peliti}. Moreover, as we have shown in Ref.
\cite{Nimw98a}, the population members in lower-fitness subbasins are
closely genetically related to members in the subbasin of currently
highest fitness. These facts flatly contradict the maximum entropy
assumption that individuals are randomly and independently spread
through the subbasins. Since these complications do not generally
alter the rate of deleterious mutations from subbasins to
lower-fitness subbasins, theoretical predictions---such as, the epoch
distributions $\vec{P}^{\vec{\imath}}$---are not much affected.
However, as shown in Ref. \cite{Nimw97b}, statistics---such as, the
average waiting time for the discovery of a portal---may be
significantly affected. This leaves open the question of how to extend
the set of macroscopic variables to account for these complications.

\medskip
\noindent
{\bf Acknowledgments.}
This work was supported at the Santa Fe Institute by grants from the
NSF, %IRI-9705830
ONR, %grant N00014-95-1-0524
and Sandia National Laboratory. %contract AU-4978.

\bibliography{epev}

\begin{thebibliography}{10}

\bibitem{Adam95a}
C.~Adami.
\newblock Self-organized criticality in living systems.
\newblock {\em Phys. Lett. A}, 203:29--32, 1995.

\bibitem{Adel94a}
L.~M. Adelman.
\newblock Molecular computation of solutions to combinatorial problems.
\newblock {\em Science}, 266:1021--1024, 1994.

\bibitem{Alte95a}
L.~Altenberg.
\newblock Genome growth and the evolution of the genotype-phenotype map.
\newblock In W.~Banzhaf and F.~H. Eeckman, editors, {\em Evolution and
  Biocomputation. Computational Models of Evolution, Monterey, California, July
  1992}, pages 205--259, Berlin, 1995. Springer Verlag.

\bibitem{Back96a}
T.~Back.
\newblock {\em Evolutionary algorithms in theory and practice: {E}volution
  strategies, evolutionary programming, genetic algorithms}.
\newblock Oxford University Press, New York, 1996.

\bibitem{Barnett97}
L.~Barnett.
\newblock Tangled webs: {E}volutionary dynamics on fitness landscapes with
  neutrality.
\newblock Master's thesis, School of Cognitive Sciences, University of East
  Sussex, Brighton, 1997.
\newblock http://www.cogs.susx.ac.uk/ lab/adapt/nnbib.html.

\bibitem{ICGA91}
R.~K. Belew and L.~B. Booker, editors.
\newblock {\em Proceedings of the Fourth International Conference on Genetic
  Algorithms}.
\newblock Morgan Kaufmann, San Mateo, CA, 1991.

\bibitem{Binn92a}
J.~J. Binney, N.~J. Dowrick, A.~J. Fisher, and M.~E.~J. Newman.
\newblock {\em The Theory of Critical Phenomena: An Introduction to the
  Renormalization Group}.
\newblock Oxford Science Publications, 1992.

\bibitem{Cham95a}
L.~Chambers, editor.
\newblock {\em Practical Handbook of Genetic Algorithms}.
\newblock CRC Press, Boca Raton, 1995.

\bibitem{Chen99a}
J.~Chen, E.~Antipov, B.~Lemieux, W.~Cedeno, and D.~H. Wood.
\newblock {DNA} computing implementing genetic algorithms.
\newblock In L.~F. Landweber, E.~Winfree, R.~Lipton, and S.~Freeland, editors,
  {\em Evolution as Computation}, pages 39--49, New York, 1999. Springer
  Verlag.

\bibitem{Crutchfield&Mitchell95a}
J.~P. Crutchfield and M.~Mitchell.
\newblock The evolution of emergent computation.
\newblock {\em Proc. Natl. Acad. Sci. U.S.A.}, 92:10742--10746, 1995.

\bibitem{Davis91a}
L.~D. Davis, editor.
\newblock {\em The Handbook of Genetic Algorithms}.
\newblock Van Nostrand Reinhold, 1991.

\bibitem{Derrida&Peliti}
B.~Derrida and L.~Peliti.
\newblock Evolution in a flat fitness landscape.
\newblock {\em Bull. Math. Bio.}, 53(3):355--382, 1991.

\bibitem{Eigen71}
M.~Eigen.
\newblock Self-organization of matter and the evolution of biological
  macromolecules.
\newblock {\em Naturwissen.}, 58:465--523, 1971.

\bibitem{Eigen&McCasKill&Schuster89}
M.~Eigen, J.~McCaskill, and P.~Schuster.
\newblock The molecular quasispecies.
\newblock {\em Adv. Chem. Phys.}, 75:149--263, 1989.

\bibitem{Elena&Cooper&Lenski96}
S.~F. Elena, V.~S. Cooper, and R.~E. Lenski.
\newblock Punctuated evolution caused by selection of rare beneficial
  mutations.
\newblock {\em Science}, 272:1802--1804, 1996.

\bibitem{ICGA95}
L.~Eshelman, editor.
\newblock {\em Proceedings of the Sixth International Conference on Genetic
  Algorithms}.
\newblock Morgan Kaufmann, San Mateo, CA, 1995.

\bibitem{Ewensbook}
W.~J. Ewens.
\newblock {\em Mathematical Population Genetics}, volume~9 of {\em
  Biomathematics}.
\newblock Springer-Verlag, 1979.

\bibitem{Font98a}
W.~Fontana and P.~Schuster.
\newblock Continuity in evolution: On the nature of transitions.
\newblock {\em Science}, 280:1451--5, 1998.

\bibitem{Font93a}
W.~Fontana, P.~F. Stadler, E.~G. Bornberg-Bauer, T.~Griesmacher, I.~L.
  Hofacker, M.~Tacker, P.~Tarazona, E.~D. Weinberger, and P.~Schuster.
\newblock {RNA} folding and combinatory landscapes.
\newblock {\em Phys. Rev. E}, 47:2083--2099, 1992.

\bibitem{ICGA93}
S.~Forrest, editor.
\newblock {\em Proceedings of the Fifth International Conference on Genetic
  Algorithms}.
\newblock Morgan Kaufmann, San Mateo, CA, 1993.

\bibitem{Forst&Reidys&Weber95}
C.~V. Forst, C.~Reidys, and J.~Weber.
\newblock Evolutionary dynamics and optimizations: Neutral networks as model
  landscapes for {RNA} secondary-structure folding landscape.
\newblock In F.~Moran, A.~Moreno, J.~Merelo, and P.~Chacon, editors, {\em
  Advances in Artificial Life}, volume 929 of {\em Lecture Notes in Artificial
  Intelligence}. Springer, 1995.
\newblock SFI preprint 95-20-094.

\bibitem{Goldberg89c}
D.~E. Goldberg.
\newblock {\em Genetic Algorithms in Search, Optimization, and Machine
  Learning}.
\newblock Addison-Wesley, Reading, MA, 1989.

\bibitem{Gould&Eldredge77}
S.~J. Gould and N.~Eldredge.
\newblock Punctuated equilibria: {T}he tempo and mode of evolution
  reconsidered.
\newblock {\em Paleobiology}, 3:115--251, 1977.

\bibitem{Hartl&Clark}
D.~L. Hartl and A.~G. Clark.
\newblock {\em Principles of population genetics}.
\newblock Sinauer Associates, second edition, 1989.

\bibitem{Haygood}
R.~Haygood.
\newblock The structure of {R}oyal {R}oad fitness epochs.
\newblock {\em Evolutionary Computation}, submitted, 1997.
\newblock ftp://ftp.itd.ucdavis.edu/pub/people/rch/ StrucRoyRdFitEp.ps.gz.

\bibitem{Huynen95}
M.~Huynen.
\newblock Exploring phenotype space through neutral evolution.
\newblock {\em J. of Mol. Evol.}, 43:165--169, 1996.

\bibitem{Huynen&Stadler&Fontana}
M.~Huynen, P.~F. Stadler, and W.~Fontana.
\newblock Smoothness within ruggedness: The role of neutrality in adaptation.
\newblock {\em Proc. Natl. Acad. Sci. USA}, 93:397--401, 1996.

\bibitem{Kauf87a}
S.~A. Kauffman and S.~Levin.
\newblock Towards a general theory of adaptive walks in rugged fitness
  landscapes.
\newblock {\em J. Theo. Bio.}, 128:11--45, 1987.

\bibitem{Kimurabook}
M.~Kimura.
\newblock {\em The Neutral Theory of Molecular Evolution}.
\newblock Cambridge University Press, 1983.

\bibitem{Koza93a}
J.~R. Koza.
\newblock {\em Genetic Programming: On the Programming of Computers by Means of
  Natural Selection}.
\newblock MIT Press, Cambridge, MA, 1992.

\bibitem{Land99a}
L.~F. Landweber and L.~Kari.
\newblock Universal molcular computation in ciliates.
\newblock In L.~F. Landweber, E.~Winfree, R.~Lipton, and S.~Freeland, editors,
  {\em Evolution as Computation}, pages 51--66, New York, 1999. Springer
  Verlag.

\bibitem{Mack89a}
C.~A. Macken and A.~S. Perelson.
\newblock Protein evolution in rugged fitness landscapes.
\newblock {\em Proc. Nat. Acad. Sci. USA}, 86:6191--6195, 1989.

\bibitem{Mitchell96a}
M.~Mitchell.
\newblock {\em An Introduction to Genetic Algorithms}.
\newblock MIT Press, Cambridge, MA, 1996.

\bibitem{MitchellEtAl93c}
M.~Mitchell, J.~P. Crutchfield, and P.~T. Hraber.
\newblock Evolving cellular automata to perform computations: Mechanisms and
  impediments.
\newblock {\em Physica D}, 75:361--391, 1994.

\bibitem{Newman&Engelhardt98}
M.~Newman and R.~Engelhardt.
\newblock Effect of neutral selection on the evolution of molecular species.
\newblock {\em Proc. R. Soc. London B.}, 256:1333--1338, 1998.

\bibitem{Nix&Vose91}
A.~E. Nix and M.~D. Vose.
\newblock Modeling genetic algorithms with {M}arkov chains.
\newblock {\em Ann. Math. Art. Intel.}, 5, 1991.

\bibitem{PrugelBennett97}
A.~Pr\"{u}gel-Bennett.
\newblock Modelling evolving populations.
\newblock {\em J. Theo. Bio.}, 185:81--95, 1997.

\bibitem{PrugelBennett&Shapiro94}
A.~Pr\"{u}gel-Bennett and J.~L. Shapiro.
\newblock Analysis of genetic algorithms using statistical mechanics.
\newblock {\em Phys. Rev. Lett.}, 72(9):1305--1309, 1994.

\bibitem{Rattray&Shapiro96}
M.~Rattray and J.~L. Shapiro.
\newblock The dynamics of a genetic algorithm for a simple learning problem.
\newblock {\em J. of Phys. A}, 29(23):7451--7473, 1996.

\bibitem{Reic80a}
L.~E. Reichl.
\newblock {\em A Modern Course in Statistical Physics}.
\newblock University of Texas, Austin, 1980.

\bibitem{Reidys98b}
C.~M. Reidys, C.~V. Forst, and P.~K. Schuster.
\newblock Replication and mutation on neutral networks of {RNA} secondary
  structures.
\newblock {\em Bull. Math. Biol.}, submitted, 1998.
\newblock SFI Working Paper 98-04-036.

\bibitem{Stre95a}
R.~F Streater.
\newblock {\em Statistical Dynamics: {A} Stochastic Approach to Nonequilibrium
  Thermodynamics}.
\newblock Imperial College Press, London, 1995.

\bibitem{Nimw98b}
E.~van Nimwegen and J.~P. Crutchfield.
\newblock Optimizing epochal evolutionary search: Population-size dependent
  theory.
\newblock {\em Machine Learning}, submitted, 1998.
\newblock Santa Fe Institute Working Paper 98-10-090. adap-org/9810004.

\bibitem{Nimw98a}
E.~van Nimwegen and J.~P. Crutchfield.
\newblock Optimizing epochal evolutionary search: Population-size independent
  theory.
\newblock {\em Computer Methods in Applied Mechanics and Engineering}, to
  appear, 1998.
\newblock Special issue on Evolutionary and Genetic Algorithms in Computational
  Mechanics and Engineering, D. Goldberg and K. Deb, editors. Santa Fe
  Institute Working Paper 98-06-046; adap-org/9810003.

\bibitem{Nimw97a}
E.~van Nimwegen, J.~P. Crutchfield, and M.~Mitchell.
\newblock Finite populations induce metastability in evolutionary search.
\newblock {\em Phys. Lett. A}, 229:144--150, 1997.

\bibitem{Nimw97b}
E.~van Nimwegen, J.~P. Crutchfield, and M.~Mitchell.
\newblock Statistical dynamics of the {R}oyal {R}oad genetic algorithm.
\newblock {\em Theoretical Computer Science}, in press, 1998.
\newblock Special issue on {E}volutionary {C}omputation, A. Eiben and G.
  Rudolph, editors. SFI working paper 97-04-35.

\bibitem{Vose93}
M.~D. Vose.
\newblock Modeling simple genetic algorithms.
\newblock In L.~D. Whitley, editor, {\em Foundations of Genetic Algorithms 2},
  San Mateo, CA, 1993. Morgan Kauffman.

\bibitem{Vose&Liepins91}
M.~D. Vose and G.~E. Liepins.
\newblock Punctuated equilibria in genetic search.
\newblock {\em Complex Systems}, 5:31--44, 1991.

\bibitem{Weber97}
J.~Weber.
\newblock {\em Dynamics of Neutral Evolution. A case study on RNA secondary
  structures}.
\newblock PhD thesis, Biologisch-Pharmazeutischen Fakult\"{a}t der Friedrich
  Schiller-Universit\"{a}t Jena, 1996.
\newblock http://www.tbi.univie.ac.at/papers/ PhD\_theses.html.

\bibitem{Wright82a}
S.~Wright.
\newblock Character change, speciation, and the higher taxa.
\newblock {\em Evolution}, 36:427--43, 1982.

\bibitem{Yeom92a}
J.~M. Yeomans.
\newblock {\em Statistical Mechanics of Phase Transitions}.
\newblock Clarendon Press, Oxford, 1992.

\end{thebibliography}
\bibliographystyle{plain}

%\end{multicols}

\end{document}